# Ultra-high-Q free space coupling to microtoroid resonators

Sartanee Suebka[1], Euan McLeod[1], and Judith Su[1,2*]
[1]Wyant College of Optical Sciences, University of Arizona, Tucson, AZ, USA
[2]Department of Biomedical Engineering, University of Arizona, Tucson, AZ, USA

## Abstract

Whispering gallery mode (WGM) microtoroid resonators are one of the most sensitive biochemical sensors in existence, capable of detecting single molecules. The main barrier for translating these devices out of the laboratory is that light is evanescently coupled into these devices though a tapered optical fiber. This hinders translation of these devices as the taper is fragile, suffers from mechanical vibration, and requires precise positioning. Here, we eliminate the need for an optical fiber by coupling light into and out from a toroid via free-space coupling and monitoring the scattered resonant light. A single long working distance objective lens combined with a digital micromirror device (DMD) was used for light injection, scattered light collection, and imaging. We obtain Q-factors as high as $1.6 \times 10^8$ with this approach. Electromagnetically induced transparency (EIT)-like and Fano resonances were observed in a single cavity due to indirect coupling in free space. This enables improved sensing sensitivity. The large effective coupling area (~10 $\mu m$ in diameter for numerical aperture = 0.14) removes the need for precise positioning. Sensing performance was verified by combining the system with the frequency locked whispering evanescent resonator (FLOWER) approach to perform temperature sensing experiments. We believe that this work will be a foundation for expanding the implementation of WGM microtoroid resonators to real-world applications.

## Introduction

Whispering gallery mode (WGM) microtoroid resonators are one of the most sensitive sensors in existence due to their long photon confinement time (~10 ns)[1] which results in a Q-factor in excess of 100 million[2]. This enables repeated interaction of light with target analytes. The high sensitivity, rapid response[3], and label-free nature of WGM resonators has enabled many biochemical applications including protein detection[4–6], drug screening[7], ovarian cancer screening[8], exosome detection,[9] and early detection of hazardous gases[1]. FLOWER (frequency locked optical whispering evanescent resonator) combines WGM devices with balanced detection and data processing[4,10] to detect attomolar protein concentrations in a time scale of seconds[3,4] and hazardous gases at part-per-trillion concentrations[11].

Light is typically evanescently coupled into microtoroid resonators through a tapered optical fiber hundreds of nanometers in diameter[4]. In spite of a high coupling efficiency in excess of 99%[12], these tapered fibers are fragile and suffer from vibrations due to fluid flow or air currents. Precise alignment of the fiber with the microtoroid is also needed for phase matching and thus efficient energy transfer[13]. The drawbacks of using a tapered fiber hinders these systems from being integrated into compact and portable lab-on-chip platforms[14] and from being multiplexed. Other coupling approaches such as prism coupling[1] are difficult to use with microtoroid-shape resonators. Microtoroids have advantages over other WGM sensors due to their on-chip fabrication as well as larger capture area which enables a faster response time compared to plasmonic sensors[15,16].

Previously, free-space coupling of light into a deformed, non-azimuthal symmetric microtoroid[17–19] due to chaos-assisted momentum transformation[20] was demonstrated; however, the irregular spectra and mode field distribution of these toroids limit their use in applications such as frequency comb generation[21,22] and evanescent biosensing, which prefer a predictable response[23]. Another approach to couple light into a microtoroid from free space has been to add nano couplers randomly positioned on the microtoroid surface for indirect coupling[24]. In that approach, a fiber lens was used to deliver free-space light; however, a tapered fiber was still used to couple the light out from the cavity and precise alignment was still required for the fiber lens.

Here, we designed a free-space coupling system for symmetric microtoroids by using a single objective lens together with a digital micromirror device (DMD). This configuration has three purposes: focusing the input light, collecting the resonant scattered light, and imaging the microtoroid. Using a single objective lens for these tasks provides a more compact system, a cheaper design, and easier alignment. A region of interest (ROI) can be selected using the DMD, which filters out some of the stray light. The advantage of imaging the resonant back-scattered light over the light transmitted in the direction of the incident beam is that there is lower background and fewer Fabry-Perot effects, which lead to higher intensity contrast[25]. An analytic expression for the free space coupling efficiency to WGM resonators exists,[26] suggesting that maximum coupling efficiency can be enhanced by reducing the size of the Gaussian beam waist. Based on this, we compared resonant scattering power between two different objective lenses: one with a numerical aperture (NA) of 0.14 and the other with an NA of 0.42. According to the analytical solution[26], a Gaussian beam can excite multiple azimuthal modes, denoted by mode number $m$. This creates a likelihood for the modes to overlap and generate EIT-like and Fano resonances. We examined the detuning between modes by changing the coupling strength via adjusting the beam-cavity distance. We created a coupling map to study free-space coupling position tolerance. Lastly, we demonstrated that our free space coupling system combined with FLOWER could be used for sensing applications by experimentally tracking changes in temperature.

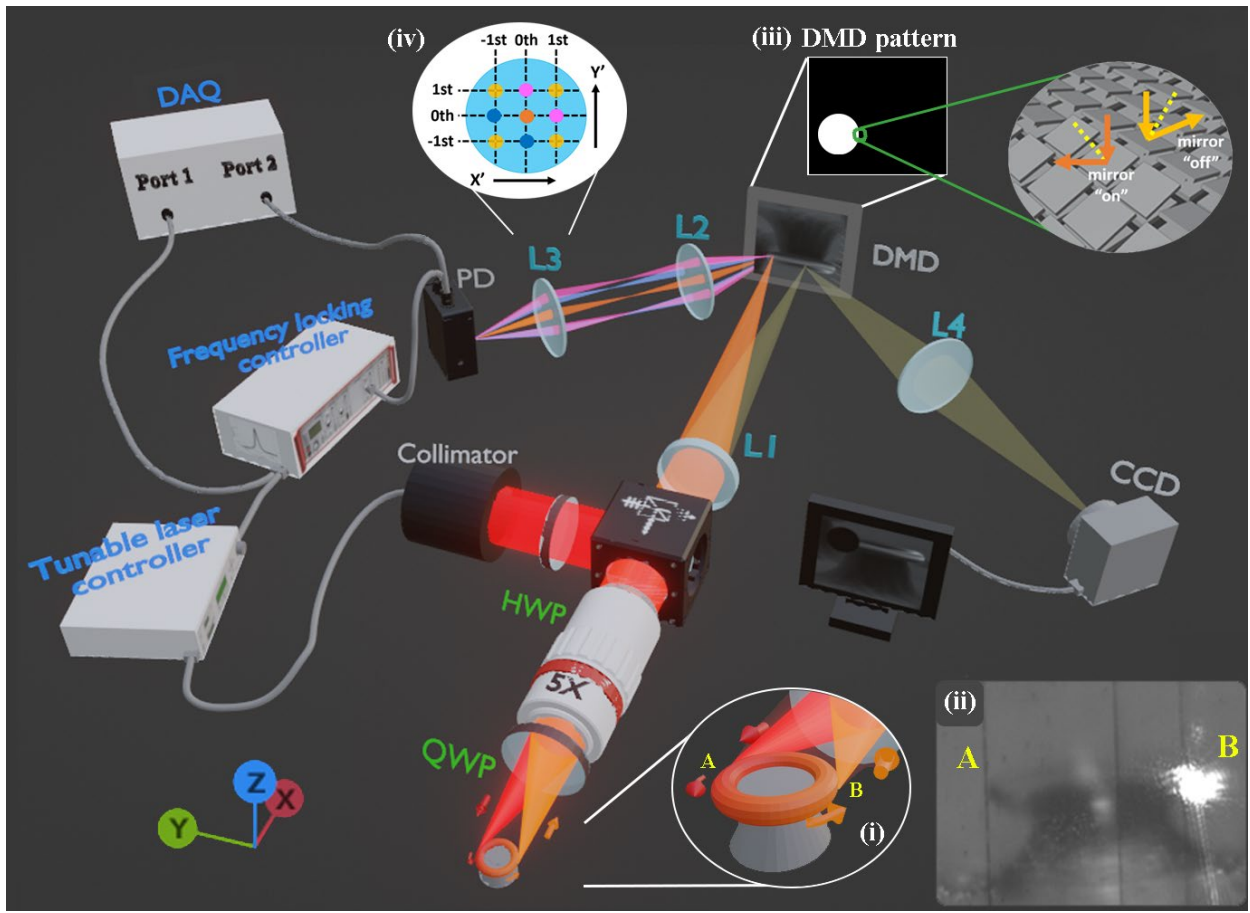

**Fig. 1** Overview of free-space coupling system. L1: tube lens. L2 and L3 are bi-convex lenses and together form a 4f-configuration to collect several diffraction orders on the photodetector (PD). L4, which is also a bi-convex lens, is the imaging lens. The yellow-brown cone indicates illumination light which comes from a ring light around the objective and not from the cavity. Inset (i) Schematic of the coupled microtoroid in free space. The laser converges at edge A to couple into the cavity as indicated by the red cone. At the resonance wavelength, the coupled wavelength is confined in the cavity. The scattering light from edge B is then collected by the same objective lens to observe the resonance wavelength as indicated by the orange cone. Inset (ii) Image from the CCD during an experiment. Vertical black lines are due to inactive pixel lines. Inset (iii) DMD pattern to select the ROI. Micromirrors in the white area are directed to the PD. The mirrors in the black area are directed to CCD for imaging. Inset (iv) Schematic of different diffraction orders on the L3 plane.

## Materials and Methods

### Microtoroid fabrication

Microtoroid resonators were fabricated as previously described[4] using photolithography and thermal $CO_2$ laser reflow. Shipley S1813 photoresist was spun coat on a 2 $\mu m$ thermally grown silica ($SiO_2$) layer on top of a silicon wafer (University Wafer, MA). After exposing UV light through a photomask consisting of opaque arrays of 150 $\mu m$ diameter circles to the photoresist layer, 150 $\mu m$ diameter circular discs were patterned on the photoresist. The exposed silica areas were etched away using a 6:1 (v/v) buffered oxide etchant. Etching stopped at the silicon layer. The remaining photoresist was removed with acetone and washed with isopropyl alcohol. Silica discs with diameter of 150 $\mu m$ were left on silicon substrate. Samples were baked at 175 °C for at least 10 mins to remove moisture. A xenon difluoride ($XeF_2$) etch was performed, undercutting the silicon and creating silica microdisks. Finally, thermal reflow by a $CO_2$ laser (Synrad, WA) was done to produce the finished microtoroid resonator structure which had a major and minor diameter of ~50 $\mu m$ and ~4 $\mu m$, respectively (see Fig. S1 in Supplementary Note 1).

### Optical configuration

An overview of the free-space coupling system is shown in Fig. 1. A tunable laser (Velocity™ TLB-6712, Newport) with a tuning range of 765-780 nm is used. A long working distance objective lens (NA = 0.14, 5× M Plan Apo NIR, Mitutoyo) is used for three purposes: focusing laser light, collecting scattered light, and imaging the microtoroid. A DMD (854x480 pixels, pixel pitch of 5.4 $\mu m$, DLP2010, Texas Instruments) separates the scattered light from the cavity into two beams: one for monitoring the resonance wavelength shift using a photodetector and the other for imaging the microtoroid. In later experiments, a higher NA objective lens (NA = 0.42, 20× M Plan Apo, Mitutoyo) was used for comparison.

The laser, indicated by the red path, passes through the collimator, half wave plate (HWP), polarizing beam spitter (PBS), objective lens, and quarter wave plate (QWP) before reaching the toroid. The HWP is used for polarization rotation to maximize the output power. The QWP changes the polarization state of the scattered light so that it transmits through the PBS. Here, we define the TE and TM waves to have polarization parallel to the z-axis and xy-plane, respectively. The light reflected light from the PBS is TE, which becomes circularly polarized after passing through the QWP. This configuration can also behave like an isolator to reduce the back reflection from the objective lens, although it is not a true isolator since there is no external magnetic field in our system. Output free-space light converges to one of the microtoroid edges (edge A shown in Fig. 1 inset (i-ii)). On-resonance wavelengths are then coupled and confined in the cavity. Some amount of the confined light in the cavity scatters out as indicated by orange path. The objective lens collects scattered light from the opposite edge (edge B shown in Fig. 1 inset (i-ii)). Because the polarization state of the out-coupled light has been modified from that of the in-coupled light due to resonance within the microtoroid, a significant amount of light can pass through the PBS rather than being reflected back to the source. This transmitted light then passes through tube lens (L1) and converges on the DMD. For imaging purposes, a ring light is fitted around the objective lens for illumination. The illumination scattered back from the microtoroid, shown by yellow-brown cones in Fig. 1, goes through the objective lens and is imaged on the DMD plane.

To distinguish the resonant scattered light from illumination light, custom LabVIEW software was developed to upload the DMD pattern and select a ROI. As shown in Fig. 1 inset (iii), micromirrors in the white region tilt to a +17° angle, which directs light to the photodetector (PD) (PDA100A2, THORLABS). After the DMD, light diffracts in several orders due to the grating effect of the DMD. Different diffraction orders are represented by different colors as shown in Fig. 1 inset (iv). To reduce diffraction loss, L2 and L3 are used to form a 4f-configuration. Different diffraction orders will converge to the same point at the PD. The aperture is large enough to capture the 2$^{nd}$ order. For imaging, micromirrors in the back region,

Fig. 1 inset (iii), tilt to a $-17°$ angle to direct the maximum intensity order to the imaging lens (L4) whose aperture filters out the other orders. Fig. 1 inset (ii) shows an image taken during an experiment. The input laser couples in at edge A. The resonant scattered light leaks from edge B.

During the experiment, a microtoroid chip is placed on a nanopositioner (P-611.3, Physik Instrumente) for alignment. The image from the CCD helps us to locate the microtoroid, position the toroid to focus the input light into it, and see the resonant scattered light area. After selecting the ROI around edge B from Fig. 1 inset (ii) to filter out stray light, as mentioned above, the DMD splits the resonant scattered light and illumination into two different paths. Light in the ROI is delivered to the PD. Consequently, the ROI appears dark on the CCD which also serves as a cross-reference to ensure that the ROI is at the desired location (see Fig. S3-Fig. S4 in Supplementary Note 2).

**Temperature sensing experiment**

The microtoroid temperature was adjusted using thermoelectric cooling (TEC) (TECF2S, THORLABS). The microtoroid chip and thermistor (GL202F9J, Littelfuse) were attached to the TEC using thermally conductive tape. A temperature controller (SLICE-QTC, Vescent Photonics) generated a temperature control feedback loop by reading out the temperature from the thermistor and controlling the TEC to either heat or cool. The resonance wavelength shift corresponding to temperature was then measured by two different methods: the scanning method and the frequency locking method (FLOWER).

For the scanning method, a DAQ (PCI-4461, National Instruments) recorded the resonance curve from the PD, port 2 in Fig. 1, while the tunable laser kept scanning through various wavelengths. The resonance wavelength at each time was then extracted from each resonance curve. For the frequency locking method, we used FLOWER, which was developed in our group for aqueous biological sensing[4,9,27,28]. In short, a frequency locking feedback controller reads the signal from the PD and sends out the voltage to a tunable laser controller to adjust the laser frequency to the cavity resonance. A DAQ reads the voltage from the frequency locking controller, port 1 in Fig. 1, which corresponds to the resonance wavelength shift.

## **Results and Discussions**

**Resonance line shapes and system efficiency**

In this work, we observed both asymmetric Fano resonances and generalized Fano resonances in free-space coupling to a single microtoroid cavity. Two modes were simultaneously excited. These two modes have distinct quality factors, unlike the simultaneous excitation of degenerate modes in mode-splitting based sensing[29–32] or exceptional point sensing[33–35]. Fig. 2(a-c) shows three different resonance line shapes observed: a Lorentzian (Fig. 2(a)), a standard Fano (Fig. 2(b)), and a generalized Fano line shape (Fig. 2(c)). The Lorentzian line shape is given as:

$$I_L(N) = \frac{A}{1+N^2} + B, \quad (1)$$

where $N = 2\frac{(v-v_0)}{\Gamma}$; A, B, $\Gamma$, $v$ and $v_0$ are the amplitude, the non-resonant background, the linewidth at half maximum, the frequency, and the resonance frequency, respectively. The standard Fano line shape is the product of interference between a single resonance mode and the continuum background and is given[36,37] by

$$I_{SF}(N) = F\frac{(N+q)^2}{1+N^2} + B, \quad (2)$$

where q and F are the asymmetry and amplitude factors, respectively. The linewidth of the standard Fano line shape is denoted by the frequency difference between the peak and dip[38].

In the case of the generalized Fano line shape, which is a product of interaction between two modes and the continuum background[39], the shape is given by:

$$I_{GF}(N_1, N_2) = B + \sum_{i=1}^{2} F_i \frac{(N_i + q_i)^2}{1 + N_i^2}. \tag{3}$$

The Q-factor for each resonance is defined by the ratio of $\nu_0$ to $\Gamma$. Fig. 2(c) shows one of the generalized Fano line shapes, where the first observed mode is a standard Fano line shape, ignoring the dip between the modes, shown in Fig. 2(c)inset, and the resulting quality factor is $Q_1 \sim 7.8 \times 10^7$. The dip is present due to the existence of the 2$^{nd}$ mode, which has $Q_2 \sim 1.6 \times 10^8$. See Fig. S5 in Supplementary Note 3 for more resonance line shapes examples.

The system percent efficiency was calculated by the following equation:

$$\% \text{ Efficiency} = \frac{P_{res}}{P_{input}} \times 100\%, \tag{4}$$

where $P_{input}$ and $P_{res}$ are respectively the input power after the QWP and resonance power which is calculated differently for different line shapes. For the Lorentzian line shape, $P_{res}$ is the difference between the peak/dip power and the baseline as shown in Fig. 2(a) or $A$ in eq. (1).

For the standard Fano line shape, $P_{res}$ is the power from the minimum dip to maximum peak as shown in Fig. 2(b). We derived $P_{res}$ for the standard Fano line shape, see Supplementary Note 3:

$$P_{res,Fano} = |F(1 + q^2)|, \tag{5}$$

For the generalized Fano line shape, it is difficult to directly determine $P_{res}$ from the plot. Here, we used parameters from the fitting equation (eq. (3)) together with eq. (5). Since there are two different modes, two $P_{res,Fano}$ values can be obtained.

The % efficiency and Q-factor of 140 resonance modes from 57 microtoroids were plotted in Fig. 2(d), see Fig. S7 in Supplementary Note 3 for more details. We found that the 0.42 NA microscope objective provided ~10 times greater coupling efficiency than the 0.14 NA objective. The sources of the loss are from the optical components, diffraction at the DMD, and the free-space coupling efficiency. In addition, only part of the resonant scattered light is collected. The low fraction of retrieved power can be compensated by using higher input power or by increasing the photodetector exposure time. Although the coupling efficiencies are less than 0.1%, they are more than sufficient to obtain strong resonance peaks with high SNR (SNR ≥ 26 dB at Q~$10^7$), which is all that is necessary for sensing experiments. Here, SNR is the ratio of $P_{res}$ to noise in RMS.

We find that free-space excitation is particularly conducive to accessing a wide range of cavity modes. This is due to the nature of the focused Gaussian beam spot, which can couple to multiple azimuthal modes [26]. Consequently, it is easy to find overlapping modes that produce EIT-like and Fano resonances. Generally, resonances in a single cavity have a Lorentzian line shape. The ability to access ubiquitous Fano resonances with a steeper slopes than Lorentzian line shapes enables improved sensitivity in sensing applications[40] such as biological sensing or harmful gas detection[41–43]. Optical modes in the microtoroid are quasi-TE or quasi-TM modes[44–46]. Therefore, the resonance line shape can be adjusted by changing the polarization of the incident light[44]. In the next section, we report how the line shape can be modified by changing the beam-cavity coupling distance.

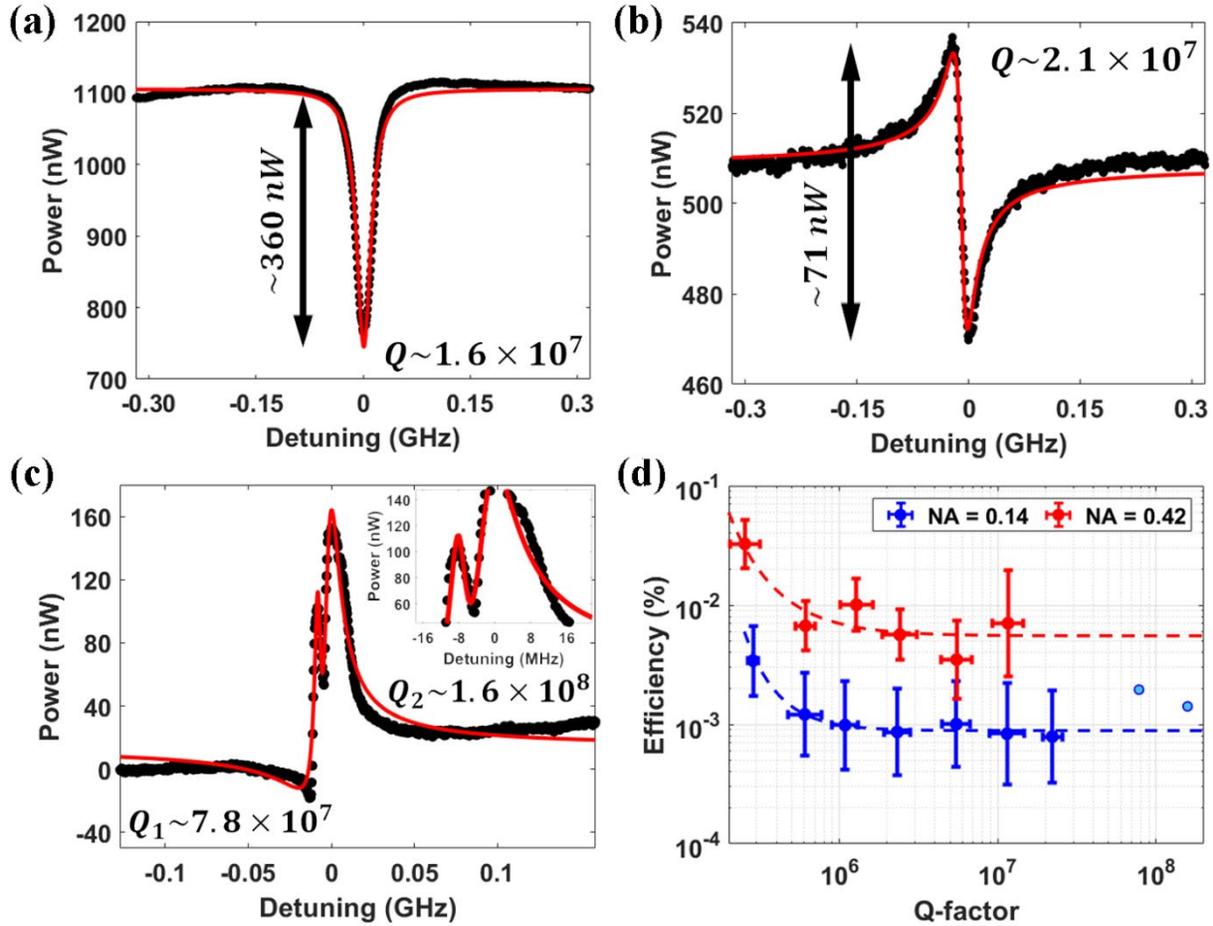

**Fig. 2** (a-c) Resonance line shapes observed from the free-space coupling system. Black dots and solid red lines show experimental results and their relevant curve fit depending on their shape. (a) Lorentzian line shape (fitted with eq. (1)) (b) standard Fano line shape (fitted with eq. (2)) (c) generalized Fano line shape (fitted with eq. (3)). (d) Efficiency vs Q-factor for two different objective lenses. Dashed lines indicate the trend. Blue and red plots show results from NA = 0.14 and 0.42 objective lenses, respectively. The data from 140 resonance modes from 57 microtoroids was divided into ten different groups by Q-factor in log scale. The error bars were then plotted as the standard deviation of the Q-factor and % efficiency in each group, see Fig. S7 in Supplementary Note 3 for more detail.

**Effect of coupling distance and coupling stability**

By changing the beam-cavity distance, the Fano line shape can be modified since a phase difference between the resonance mode and continuum mode is introduced. To explore this, we performed experiments in which the laser spot position was fixed and only the microtoroid was moved via a nanopositioner. Fig. 3(a) and (b) show how the resonance line shapes change when the beam-cavity distance changes by moving the microtoroid along the y-axis. At position 1, the resonance curve closely resembles a Lorentzian dip line shape. At later positions, Fano line shapes with different asymmetric profiles are observed. More line shapes are shown in Supplementary Movie S1. The Fano parameter $q$, which characterizes the asymmetry of the Fano profile and relates to the coupling strength between resonance modes and the continuum state, is the cotangent of the phase shift between two modes,[36] which varies with the beam-cavity distance, as shown in Fig. 3(c). In agreement with the analytic solution for circular cavities[26], the phase shift can be varied greatly in the over coupling regime ($y > 0\ \mu m$), while there is only small perturbation in phase in the under coupling regime ($y < 0\ \mu m$).

As shown in Fig. 3(d), maximum Q-factor is in the over coupling regime, not at the critical coupling condition. There is a tradeoff between resonant power and Q-factor near this range. To quantify this tradeoff, we defined the figure of merit (FoM) as[47]

$$FoM = Q \times P_{res}, \tag{6}$$

where Q is quality factor. The FoM for different beam-cavity distances is shown as a dashed blue line in Fig. 3(d). The best FoM lies in the range of $y \in [-2.5, 7.5]\ \mu m$. More resonance line shape transitions as a function of the beam-cavity distance are shown in Supplementary Note 4 (Fig. S8-Fig. S9).

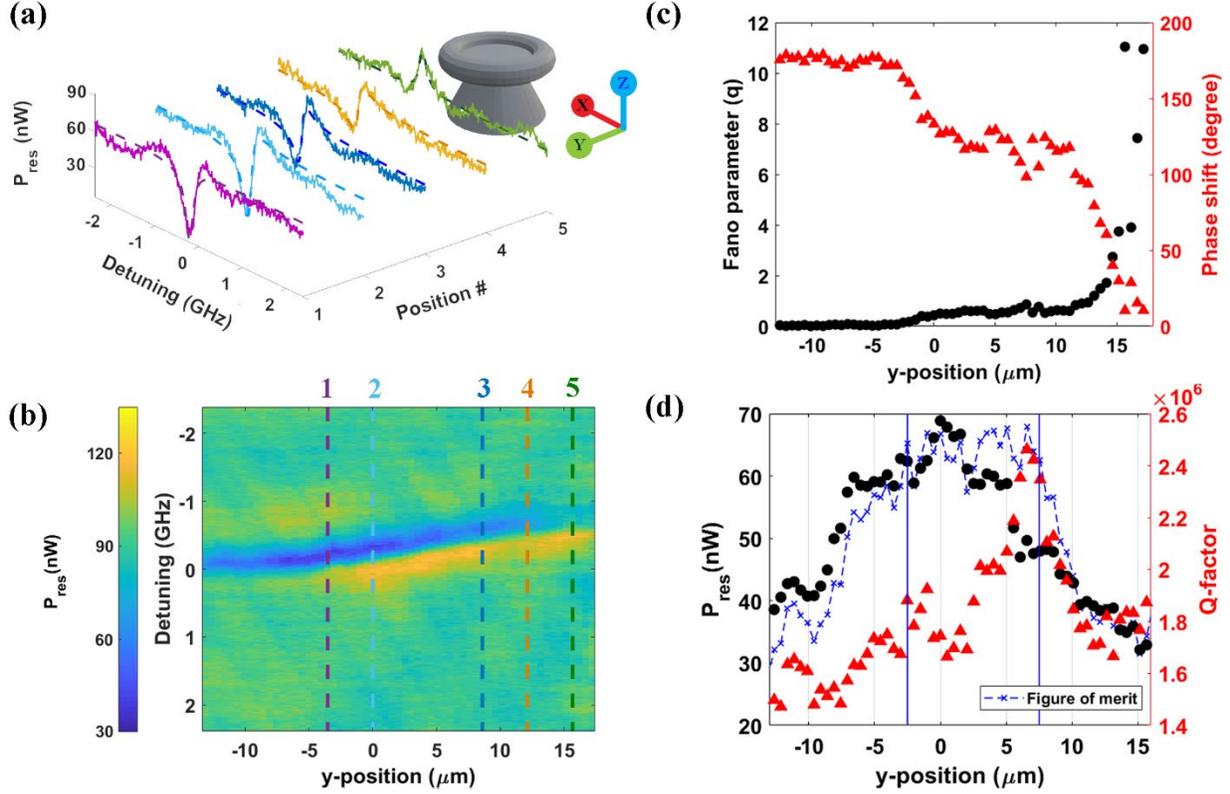

**Fig. 3** Free-space coupling map. The 5x objective lens (NA=0.14) was used for all panels. (a) Resonance curve at different microtoroid position numbers, which correspond to the lines indicated in (b). (b) Color indicates the resonance power. The y-position is defined to be zero at the highest resonance power calculated from eq. (5). Positive y-positions mean decreased beam-cavity distance. (c) Fano parameter and phase shift vs y-position. (d) Resonance power calculated from eq. (5) and Q-factor calculated from linewidth obtained from eq. (2) vs y-position. Dashed blue line indicates the figure of merit level (a.u.). y-positions in the range of -2.5 to 7.5 $\mu m$ provide maximum figure of merit.

The possibility of losing coupling by minor tapered fiber movements and mechanical vibration is one of the most problematic issues for the conventional tapered fiber coupling system. Here, we investigated the coupling area and how the position of the toroid affects the resonance by scanning the microtoroid position in 2 dimensions. Fig. 4(a) is an SEM image of the microtoroid used for scanning with a major diameter ($D_{major}$) of ~100 μm and minor diameter ($D_{minor}$) of ~8 μm. The input laser was focused on the left-hand edge of the microtoroid as shown in Fig. 4(a). The resonance curve is shown in Fig. 4(b).

First, the microtoroid is scanned along the y-axis and z-axis (Fig. 4(c-d)). Values in the +y-axis direction mean a smaller beam-cavity distance. Values in the +z-axis direction mean the microtoroid was moved upward or that the input beam moved downward relative to the microtoroid. The unfiltered detuning wavelength vs. position maps are shown in Fig. S10. As shown in Fig. S10 and Fig. 4(b), there is a background offset in the plot. We filter the background signal at each position and replot the coupling maps in Fig. 4(c-d). Similar to what was observed in fiber coupling systems[48–50], the resonance wavelength is sensitive to the gap size, especially in the over coupling regime, $y > 0\ \mu m$. It agrees with the analytic result that the resonance wavelength is detuned in the under and over coupling regimes[26]. In short, changing the coupling position can alter the coupling strength, which also modifies the effective index of the cavity mode[45].

The microtoroid was then moved in the YZ plane to generate a 2D map. Each resonance curve from a different position was fitted with a Lorentzian equation (eq. (1)). Each point in the coupling power map in Fig. 4(e) is the power from the baseline to the resonance peak, which is parameter $A$ from eq. (1). The position that has the highest power was defined to be the origin $(y, z) = (0,0)$. The resonance wavelength shifts as a function of microtoroid position are shown Fig. 4(f). By changing the position, the maximum possible resonance wavelength shift is about 1 pm. The background power, parameter $B$ from eq. (1), shown in Fig. 4(g), is the combination of stray light and power from a low Q-mode that is unresolved in this scan range. The powers along the $y$-axis at $z = 0$ and the $z$-axis at $y = 0$ are plotted in Fig. 4(h-i). The full-width at half maximum (FWHM) of the coupling power ranges from $7.1 - 11.3$ µm. If mechanical vibration amplitudes are significantly smaller than the FWHM, then they are not expected to significantly affect the sensor signal.

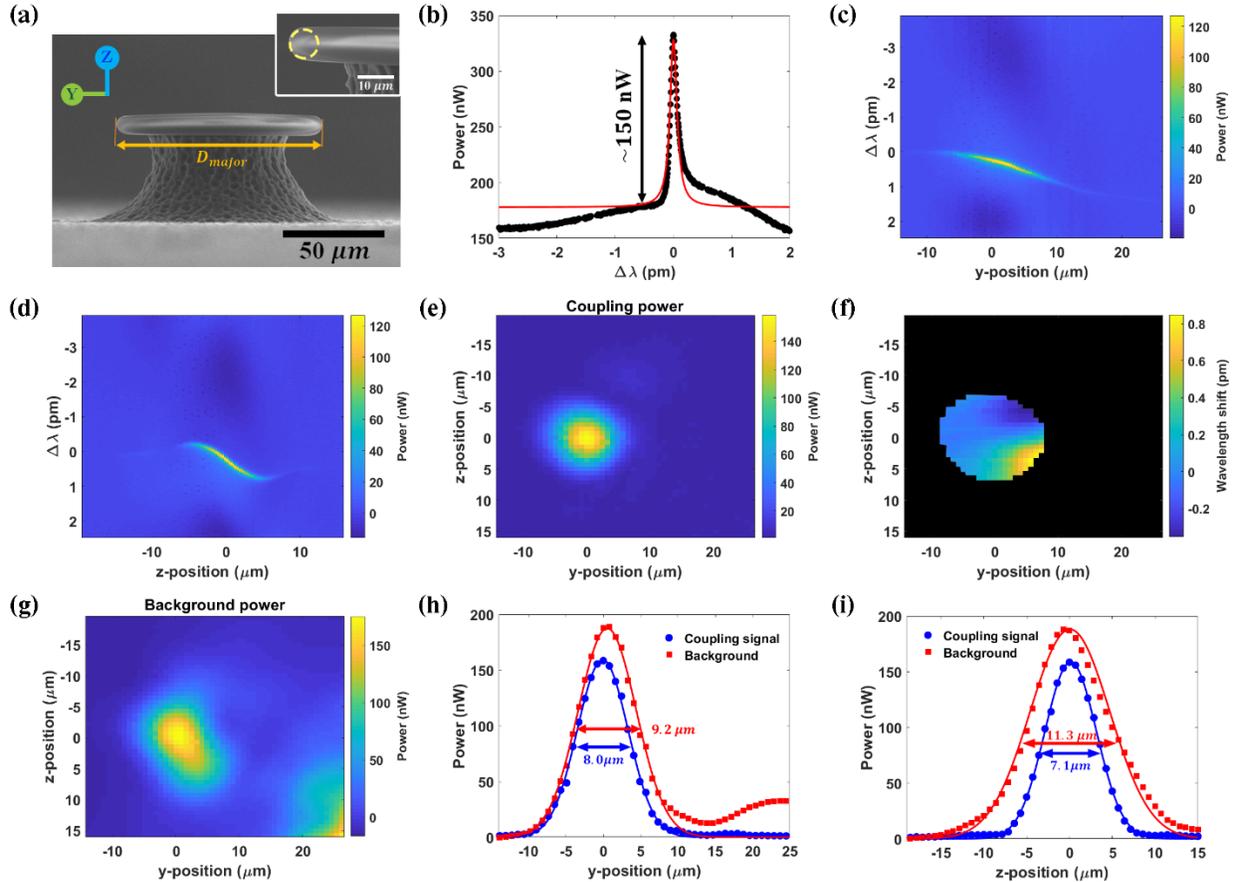

**Fig. 4** Free-space coupling map. The 5x objective lens (NA=0.14) was used for all panels. (a) SEM image of a microtoroid. The yellow circle diameter is the minor diameter. (b) Resonance curve at $(y, z) = (0,0)$, which is the position with the highest power. (c, d) Spectrograms of light scattering out of the microtoroid, when scanned along the $y$-axis (c) and the $z$-axis (d). $\Delta\lambda$ represents the detuning wavelength. (e) Resonance power map in the YZ plane (f) Resonance wavelength shift map in the YZ plane. Only data with coupling powers higher than 0.06 times the max coupling power are plotted in color. (g) Background map in the YZ plane. (h, i) Power along the $y$-axis at z = 0 (h) and along the $z$-axis at y =0 (i). Solid lines show the fits to a Gaussian equation for determining the FWHM.

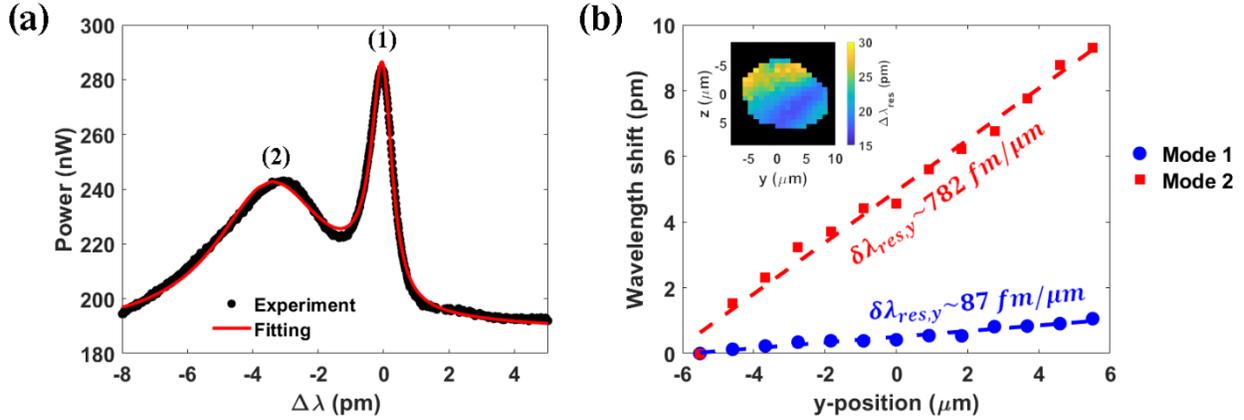

**Fig. 5** Mode crossing induced by positioning the microtoroid relative to the laser focus. (a) Resonance curve with two resonance mode fitted with eq. (3). (b) Resonance wavelength shift along y-axis at z = 0. (Inset) Map of the difference in resonance wavelength shift between the two modes.

Changing the coupling strength between the beam and cavity by adjusting their separation causes different resonance wavelength detuning for the different modes as shown in Fig. 5. Two-dimensional scanning in the YZ plane was performed to map the coupling power of two modes (Fig. S11 in Supplementary Note 4). The maximum power position for mode 1 was defined to be at (0,0). The resonance curve at (0,0) was fitted with eq. (3) in Fig. 5(a) and wavelength shifts were defined as zero. Both resonance modes were then tracked at each position in the YZ plane to create two resonance wavelength shift maps ($\Delta\lambda_{res}$). The map of the difference of the two wavelength shifts is shown in Fig. 5(b), inset. The resonance wavelength shifts along the $y$-axis at $z = 0$ are plotted in Fig. 5(b), which were fit to linear equations. When changing the microtoroid position, the wavelength of mode 2 shifted faster than that of mode 1. This leads to the possibility of modifying the line shape by adjusting the coupling position to get sharp Fano line shape or EIT-like resonances.

Using a 20× objective lens provides higher efficiency but a smaller coupling area due to a smaller spot size than that from the 5× objective lens. Two coupling maps from two different modes are shown in Fig. 6(a-b). The smaller spot size provides a higher resolution. A coupling map can also readily monitor the field distribution in the cavity. This type of mapping has been previously proposed using a tapered fiber[51]. Scanning in free-space has the advantage that it is easier to precisely control the beam-cavity position shift since perturbative electrostatic forces between the tapered fiber and the cavity are avoided. The coupling map in Fig. 6(a) shows the fundamental mode whose electric field is distributed around the equator (Fig. 6(c)). The coupling map in Fig. 6(b) has a multi-lobed distribution. The dashed white circle represents the microtoroid cross-section corresponding to the circle in Fig. 4 (a), inset. The microtoroid pillar is to the right-hand side of the map. We note that the circle was drawn as a guideline without an exact position measurement. The multiple lobes distributed around the microtoroid ring imply a higher order mode whose electric field has multiple lobes in the cavity (Fig. 6(c)). A finite element simulation was performed to show the electric field distribution in a cavity as a reference (Fig. 6(c), more details in Supplementary Note 5).

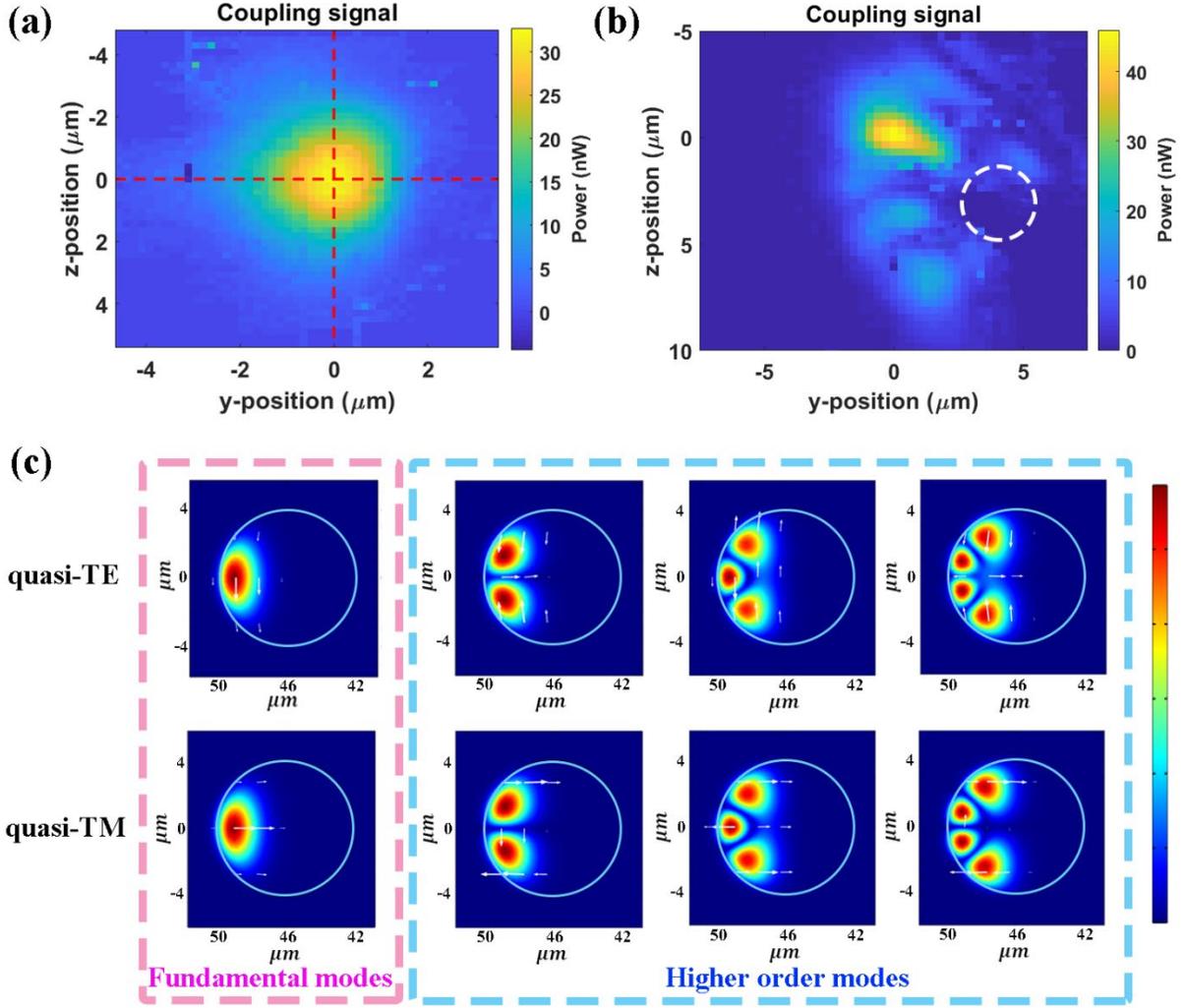

**Fig. 6** Free-space coupling map of fundamental and higher-order modes. Data was acquired using a 20× objective lens (NA = 0.42). (a) Coupling map of the fundamental mode. Dashed red lines indicate the y and z axes. (b) Coupling map of a higher order mode. The dashed white circle shows the microtoroid cross-section corresponding to the circle in Fig. 4(a), inset. The microtoroid center is on the right side. (c) Simulated electric field distribution using COMSOL. The labels along the lower axis show the distance from microtoroid's axis of revolution. The color indicates the electric field magnitude. White arrows represent the electric field direction.

Comparisons of the coupling area FWHM between a 5× and a 20× objective lens, NA = 0.14 and 0.42, respectively, are shown in Fig. 7 and Table 1. The transverse coupling area FWHM is related to the beam spot size. The spot size after a Gaussian beam is focused by an objective lens is given by:

$$d = \left(\frac{4\lambda}{\pi}\right)\left(\frac{f_{obj}}{D}\right), \quad (7)$$

where $d$, $f_{obj}$ and D are the beam waist diameter at the entrance aperture, the effective focal length of objective lens, and the input beam diameter, respectively. This equation is a paraxial approximation, which is accurate for NA < 0.9[52]. To calculate $D$, we first consider the diameter of our free-space beam, which is generated by coupling light from a single mode fiber (780-HP, Thorlabs) using a fiber bulkhead adapter and 4 cm focal length collimator lens. The output beam divergence from the single mode optical fiber ($\theta_{SM}$) is given by[53]:

$$\theta_{SM} \approx \frac{0.64\lambda}{MFD}, \tag{8}$$

where MFD is mode field diameter. $\theta_{SM}$ is half the full angular extent of the beam in radians. The collimated beam diameter ($D_c$) was then calculated by:

$$D_c = 2f_c \tan\theta_{SM} \approx 0.87\ cm, \tag{9}$$

where $f_c$ is collimator lens focal length. However, $D$ is also limited by the pupil diameter ($D_p$).

For the 5× objective lens, $D = D_c$ in eq. (7), as the pupil diameter (1.12 cm) is larger than collimated beam (0.87 cm). For the 20× objective lens, the input collimated beam overfilled the pupil (0.84 cm in diameter). Since the overfill factor is small, the spot size is approximately calculated by eq. (7) when $D = D_p$.

For the 5× objective lens, the FWHM along the $y$- and $z$-axes are $10.9 \pm 2.5\ \mu m$ and $9.4 \pm 1.9\ \mu m$. The spot size is $4.55\ \mu m$. For the 20× objective lens, the FWHM along the $y$- and $z$-axes are $2.7\ \mu m$ and $3.8\ \mu m$, respectively. The spot size is $1.18\ \mu m$. The transverse FWHM is about 2 to 3 times bigger than the spot size.

The FWHM along the $x$-axis, which is parallel to the objective lens optical axis, is 74.3 and 10.9 μm for the 5× and 20× objective lenses, respectively. The depth of focus (DOF) is given by:

$$\text{DOF} = \frac{\pi d^2}{2\lambda}, \tag{10}$$

where $d$ is the focused beam spot diameter. Therefore, the estimated DOF is 41.6 μm and 2.8 μm for the 5× and 20× objective lens, respectively. In short, higher NA provides higher coupling efficiency, but a smaller coupling area due to a smaller spot size and short DOF. The summary of coupling area FWHM, spot size and DOF comparison is shown in Table 1.

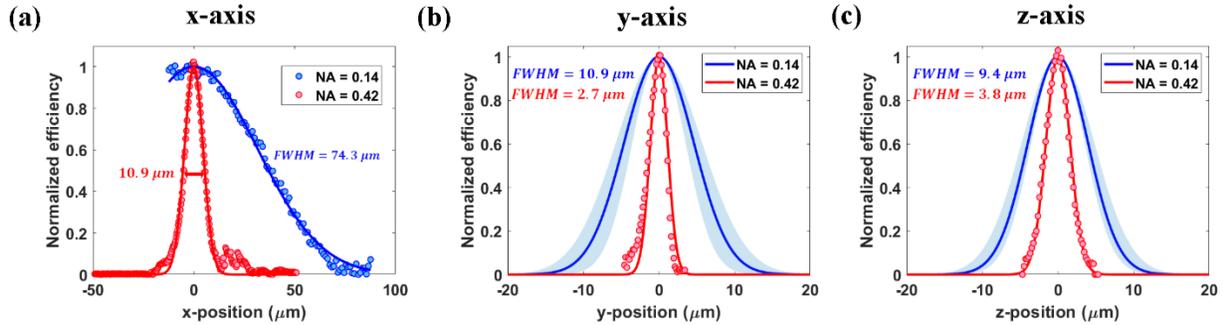

**Fig. 7** Coupling zone sizes. The normalized efficiency is shown along the (a) $x$-axis, (b) $y$-axis, and (c) $z$-axis for a 5× and a 20× objective lens (NA = 0.14 and NA = 0.42). Each curve is fitted with a Gaussian equation as shown as a solid line. For NA = 0.14 in (b-c), there are three sets of data. The blue solid lines show the fitted data to the average data. The shaded areas indicate the standard deviation from the average of experiment data (n = 3)

### Temperature sensing experiment

To verify the application of our free-space coupling system to sensing tasks, we used the microtoroid to track small changes in temperature, using the setup shown in Fig. 8(a) (see also Materials and Methods). Resonance wavelength shift corresponds to temperature. A thermistor was used to independently verify the temperature. Fig. 8(b) shows how the resonance wavelength increases with temperature. Comparisons of sensorgrams from the microtoroid and the thermistor, plotted in Fig. 8(c-d), show good agreement, confirming that our free-space coupling system can be used for sensing applications.

When tracking the resonance wavelength shift using the scanning method, the position of the resonance peak was located after each scan. There is a tradeoff between data acquisition time and resolution: shorter scanning periods lower the wavelength resolution. Also, there is another trade-off between dynamic range and resolution. FLOWER is more efficient than scanning in terms of sampling frequency and resolution since a full scan sweep is unnecessary.

The scanning method's resolution ($\delta\lambda_{res,scan}$) can be calculated by the following equation:

$$\delta\lambda_{res,scan} = \frac{2f_{sr}}{f_{PD}}(\Delta\Lambda) \tag{11}$$

where $\Delta\Lambda$, $f_{sr}$ and $f_{PD}$ are the scanning wavelength range, the scan rate, and the photodetector bandwidth. Here, the system is limited by the photodetector bandwidth (3kHz) since it is slower than DAQ bandwidth (~200kHz). A scan rate of 10 Hz and a scan range of 7 pm were used for the experiment shown in Fig. 8(c). We found $\delta\lambda_{res,scan}$ ~46 $fm$ and the scanning method's sampling frequency is 10 Hz.

Here, the tunable laser configuration is a Littman-Metcalf. The resolution is limited by how finely the piezo transducer can tune the mirror angle, which depends on the piezo voltage resolution. By using the FLOWER system, it is possible to detect wavelength shifts ($\delta\lambda_{res,FLOWER}$) at sub-attometer levels[28]. The modulation frequency (FLOWER sampling frequency) used in Fig. 8(d) was 1 kHz. The dynamic range was limited by laser frequency modulation range (60 GHz), which is equivalent to 120 pm at 775 nm. From this analysis, we can conclude that the FLOWER technique is superior to the scanning method in terms of resolution, sampling frequency, and dynamic range. Also, data analysis is simpler with the FLOWER technique.

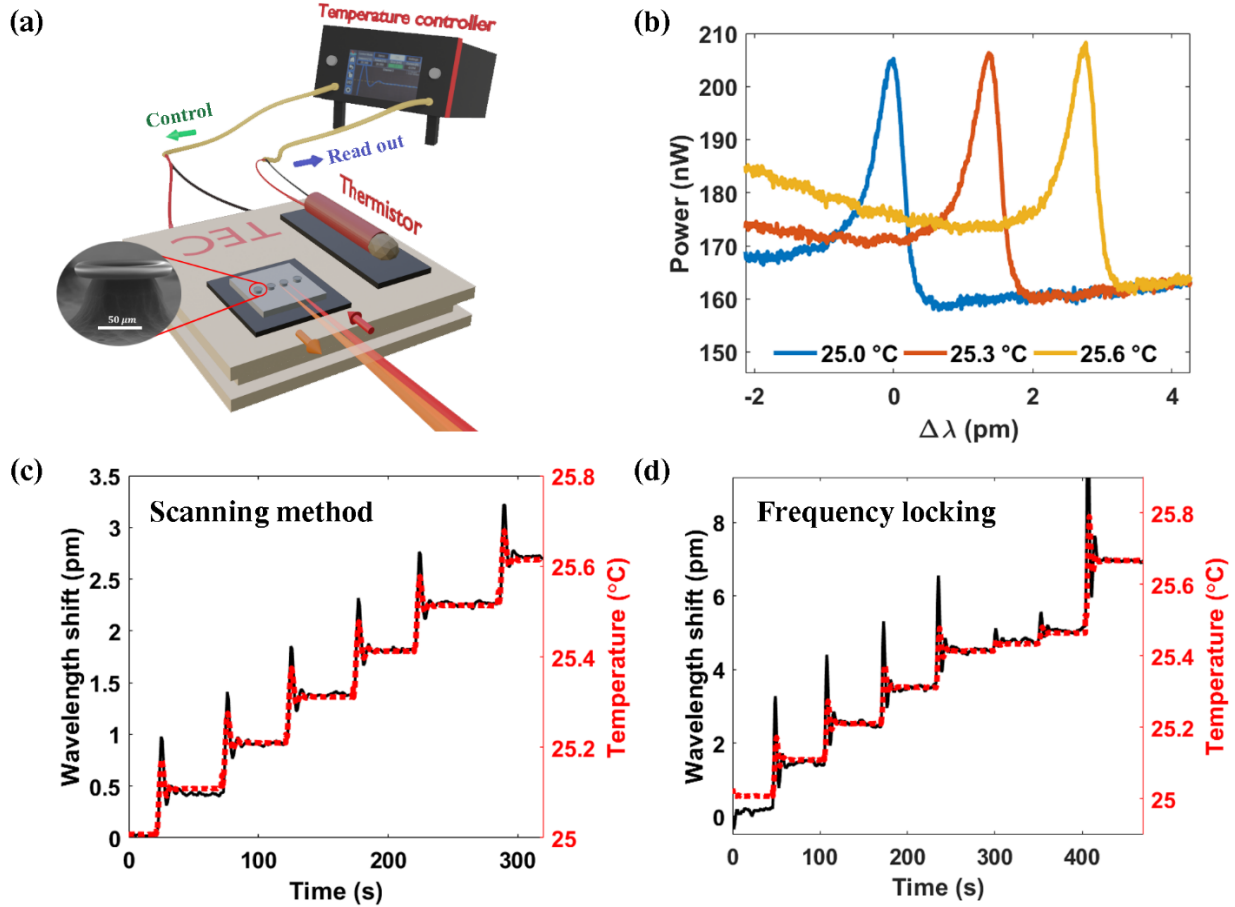

**Fig. 8** Temperature sensing experiment using free-space coupling. (a) Schematic of the experimental setup. Red and orange cones represent input and scattering light, respectively. (b) Resonance curves at difference temperatures. $\Delta\lambda$, the wavelength detuning, is defined to be zero for the peak wavelength at the starting time. (c-d) Temperature sensing results using (c) the scanning method and (d) FLOWER. The black plot is the microtoroid resonance . The red plot is the temperature measured from the thermistor.

**Conclusions**

The use of a fragile tapered fiber is a main barrier to commercializing microtoroid resonator sensors for field use. As an alternative, we introduced free-space coupling to WGM microtoroid resonators using a single long-distance objective lens by monitoring resonant scattering from the cavity. Q-factors in excess of 100 million were obtained. We showed that a more tightly focused beam can increase the free-space coupling efficiency in agreement with previous theorical predictions[26]. EIT-like and Fano resonances in a single cavity were easy to find due to multimode coupling from a Gaussian beam. This suggests that indirect coupling can be induced through free-space coupling. The Fano line shape can be modified by adjusting the beam-cavity distance because the effective refractive indices of modes shift at different rates. The sharp Fano line shape has great potential for optical switching and enhancing biochemical sensing sensitivity[44]. We introduced a FoM to quantify the tradeoff between resonant power and Q-factor. A loosely focused beam provides lower free-space coupling efficiency, but less sensitivity to mechanical vibration due to a larger effective coupling area (~10 μm in diameter for NA = 0.14). A coupling map created by scanning the beam-cavity position can be used to monitor the electric field distribution in the cavity, which also was studied by using finite element simulation. Sensing applications were verified by combining the free-space coupling system to a frequency locking technique called FLOWER. We believe that free-space coupling into microtoroid resonators can be used for spectroscopy and biosensing, and can become the foundation of fully on-chip WGM microtoroid resonator sensing systems or applications where optical fiber usage is infeasible, for example, in measuring the Q-factor of intracellular WGM lasers or where one does not have access to an optical spectrum analyzer.

**Acknowledgements**

This work was funded by the Defense Threat Reduction Agency (HDTRA1-18-1-0044).

**Conflict of interest**

JS owns a financial stake in Femtorays Technologies which develops label-free molecular sensors.

Table 1: Summary of FWHM of coupling area for 5× and 20× objective lens. Beam spot size and depth of field (DOF) are shown for comparison.

|  | Transverse FWHM ($\mu m$) | | Spot size ($\mu m$) | Longitudinal FWHM ($\mu m$) | DOF ($\mu m$) |
| --- | --- | --- | --- | --- | --- |
|  | y-axis | z-axis |  | x-axis |  |
| 5× objective lens | 10.9 ± 2.5 | 9.4 ± 1.9 | 4.55 | 74.3 | 41.62 |
| 20× objective lens | 2.7 | 3.8 | 1.18 | 10.9 | 2.8 |

# Supplementary Information for

# Ultra-high-Q free space coupling to microtoroid resonators

**This PDF file includes:**
    SEM images of microtoroids and dimensions
    Optical configuration
    Additional resonance curves and system efficiency calculation
    Additional coupling map
    Finite element simulation

**Other Supporting Online Material for this manuscript includes the following:**
    Movie S1

**MovieS1:**

**Title: Line shape transition as scanning beam-cavity distance changes**

**Caption:** Line shape transition vs different beam-cavity distance along the y-axis. Zero is defined as the position that provides maximum power.

**Supplementary Note 1:** SEM images of microtoroids and dimensions

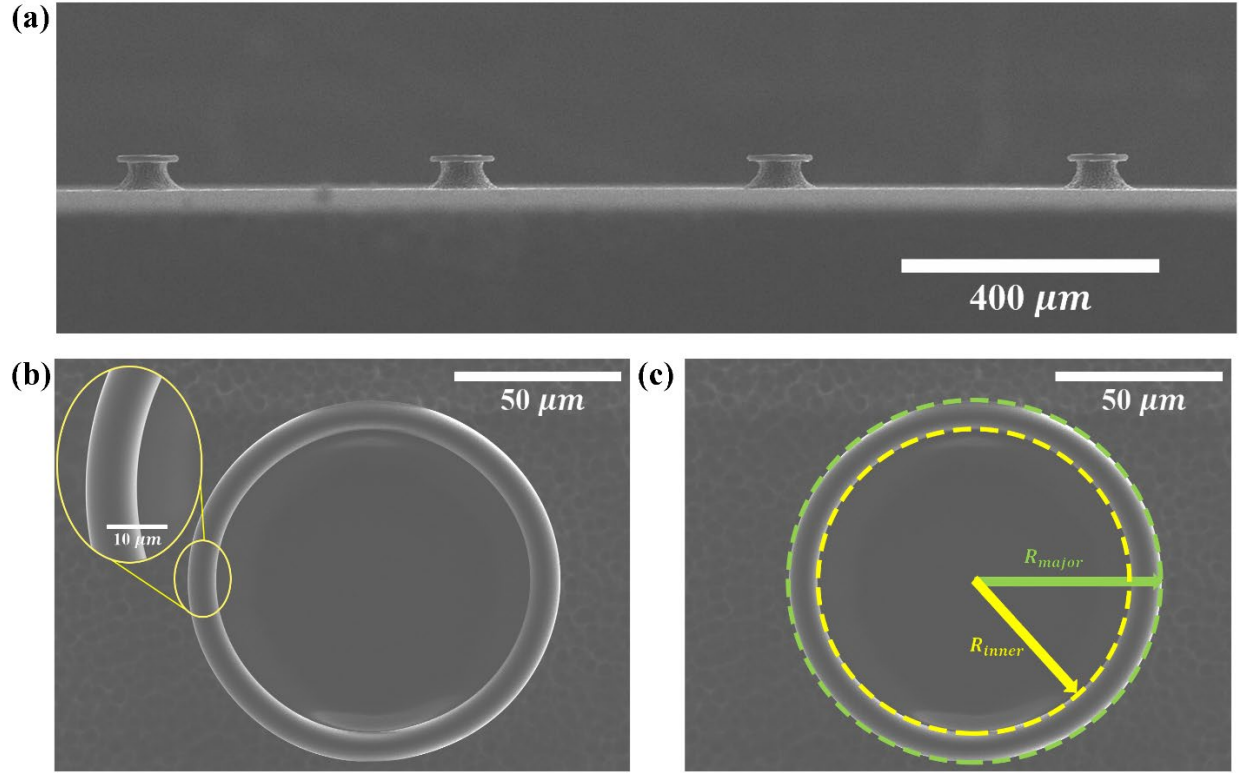

**Fig. S1** SEM images of microtoroid resonators. **(a)** Microtoroid array **(b-c)** Microtoroid; top view. Dashed yellow and green circles are drawn to find the major radius and minor radius.

The fabricated microtoroids' major radii ($R_{major}$) and minor radii ($R_{minor}$) were extracted by drawing the inner circle ($R_{inner}$) and outer circle ($R_{outer}$) as shown in Fig. S1(c). $R_{major}$ and $R_{major}$ are given by:

$$R_{major} = R_{outer} \tag{S1}$$

$$R_{minor} = \frac{R_{outer} - R_{inner}}{2} \tag{S2}$$

In these experiments, $R_{major} \sim 50$ μm and $R_{minor} \sim 4$ μm.

**Supplementary Note 2:** Optical configuration

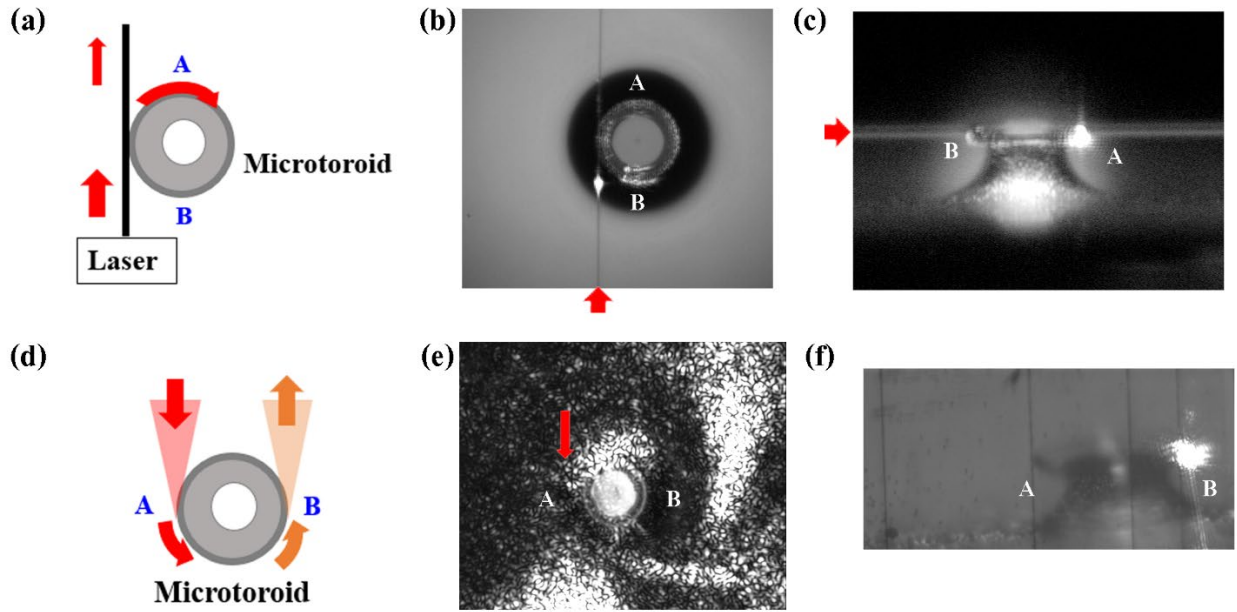

**Fig. S2** Microtoroid coupling images. **(a)** Tapered fiber coupling configuration **(b)** Microtoroid coupled with a tapered fiber, top view. The red arrow indicates the input laser. **(c)** side view of (b) **(d)** Free-space coupling configuration **(e)** Microtoroid coupled with free space light, top view. The red arrow indicates the input laser. **(f)** side view of (e) using the system in Fig. 1.

In the tapered fiber coupling system, the resonance curve is monitored through the transmitted light of the tapered fiber. It is also possible to monitor through the resonant light scattered[1] from edge A as shown in Fig. S2(c). In this work, we instead used the configuration in Fig. S2(d). By focusing light at one edge (edge A), light at resonance couples into and circulates in the cavity. Light in the cavity scattered at the other edge (edge B) is collected by an objective lens for monitoring. Fig. S2(e) shows a top view of the free space coupling experiment. Light leaking and scattering out from the cavity is circulating in the cavity as seen as a vortex-like shape in Fig. S2(e).

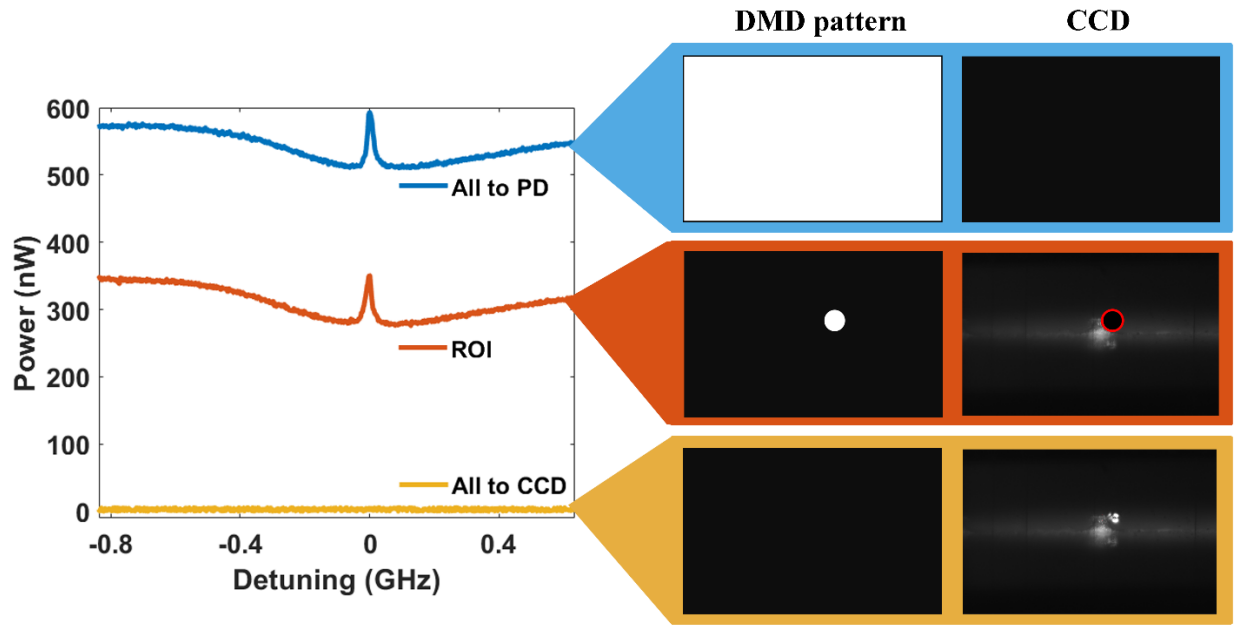

**Fig. S3** Resonance curve (left) and images from CCD (right) corresponding to DMD patterns using a 5× objective lens. The white area on the DMD indicates the area that micromirror tilts +17° to direct light to the PD. The micromirrors in the black area tilt +17° to direct light to CCD.

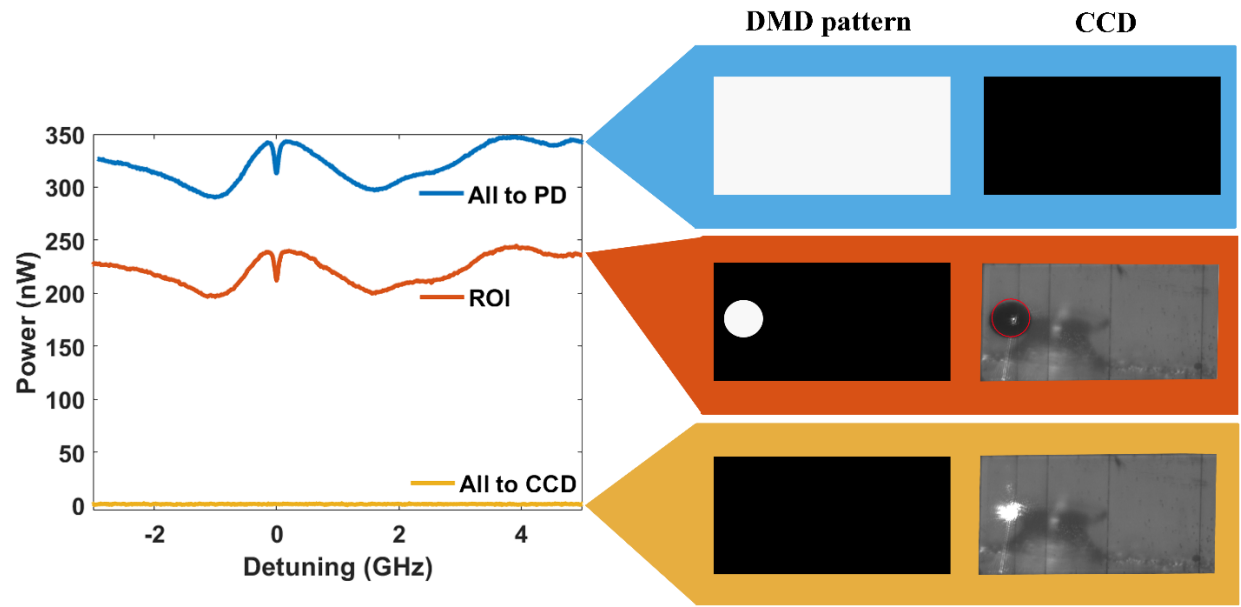

**Fig. S4** Resonance curve (left) and images from CCD (right) corresponding to DMD patterns using a 20× objective lens. The white area on the DMD indicates the area that micromirror tilts +17° to direct light to the PD. The micromirrors in the black area tilt +17° to direct light to the CCD.

In addition to the configuration in Fig. 1, a digital micromirror device (DMD) is used to select a region of interest (ROI) as shown in Fig. S3-Fig. S4. Black and white areas are used to separate the micromirror tilt angle. The micromirrors in the white and black areas direct incident light to the photodetector (PD) and CCD, respectively. By selecting an ROI, some amount of stray light is filtered out. The image on the CCD is utilized to verify if DMD pixels we selected corresponding to the ROI we want. The image appears dark at the areas tilted towards the PD.

**Supplementary Note 3:** Additional resonance curves and system efficiency calculation

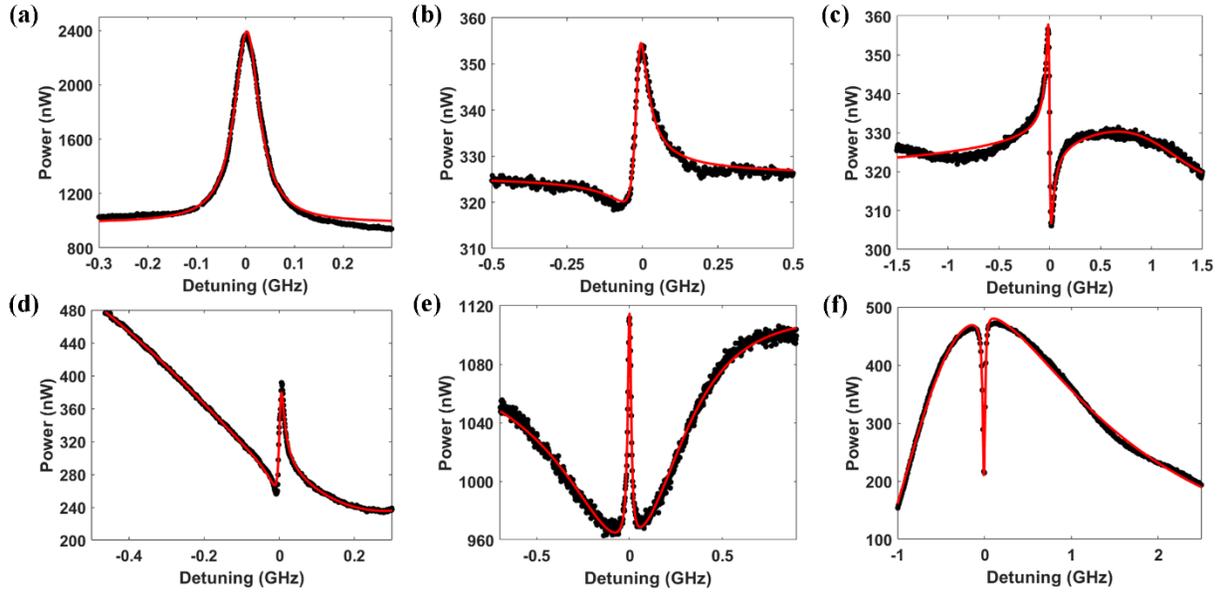

**Fig. S5** Additional resonance curves. The black dots and red lines show the experimental result and fits, respectively. **(a)** Lorentzian line shape fitted with Eq. (1) in main text. **(b)** Standard Fano line shape fitted with Eq. (2) **(c-f)** Generalized Fano line shapes fitted with Eq. (3).

Fig. S5 shows additional resonance curves beyond those already in Fig. 2(a-c). We categorized resonance curves into three types: Lorentzian, standard Fano, and generalized Fano line shapes. The Lorentzian line shape is described by Eq. (1) in main text.

In order to calculate system efficiency by Eq. (4), $P_{res}$ for the Lorentzian line shape is the power from the baseline to the peak/dip, which corresponds to the magnitude of $A$ in Eq. (1).

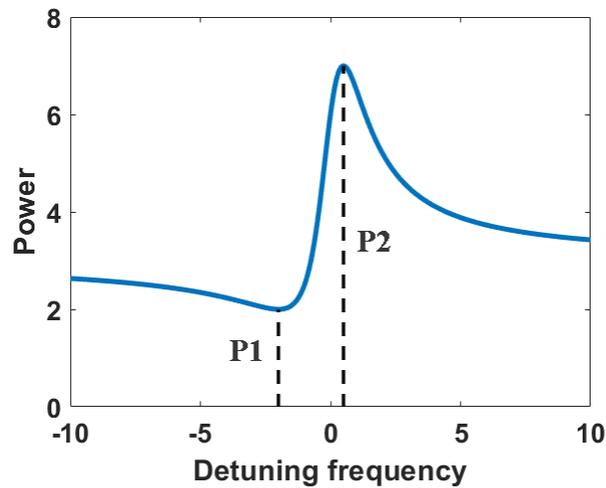

**Fig. S6** Fano profile using the formula in Eq.(2), where $F = 1$, $B = 2$, and $q = 2$. P1 is the power at the minimum. which is located at $N = -q$. P2 is the power at the maximum, which is located at $N = 1/q$.

For the standard Fano line shape (Eq. (2) in the main text), $P_{res}$ in Eq. (2) is the difference between the maximum and minimum power, $P_{res} = |P2 - P1|$, as illustrated in Fig. S6. We computed $v$ at the minimum and maximum points by taking the 1$^{st}$ derivative of Eq. (2):

$$\frac{dI_{SF}}{dN} = \frac{2F(N+q)}{1+N^2}\left[1 - \frac{N(N+q)}{1+N^2}\right] \tag{S3}$$

The 1$^{st}$ derivative is zero at the critical points.

$$0 = (N+q)[(1+N^2) - N(N+q)] \tag{S4}$$

The two solutions from Eq. (S4) are $N = -q$ and $N = 1/q$. P1 is the power where N = -q. Substituting this solution into Eq. (2),

$$P1 = B. \tag{S5}$$

Substituting $N = 1/q$ into Eq. (2),

$$P2 = \frac{F\left(\frac{1}{q}+q\right)^2}{1+\frac{1}{q^2}} + B = F(1+q^2) + B. \tag{S6}$$

$P_{res}$ is then derived:

$$P_{res} = |F(1+q^2)|. \tag{S7}$$

The generalized Fano line shape, which is a product of the interaction between two modes and a continuum background[2], is given by:

$$I_{GF}(N_1, N_2) = B + I_{1,SF}(N_1) + I_{2,SF}(N_2) = B + \sum_{i=1}^{2} F_i \frac{(N_i + q_i)^2}{1+N_i^2}, \tag{S8}$$

where $I_{1,SF}(N)$ and $I_{2,SF}(N)$ are two single Fano resonances described in Eq. (2). This is the derivation of Eq. (3) in the main text.

The fitted parameters from Eq. (S8) are substituted into Eq. (S7) to calculate $P_{res}$ for each mode as given by:

$$P_{res,i} = |F_i(1+q_i^2)| \tag{S9}$$

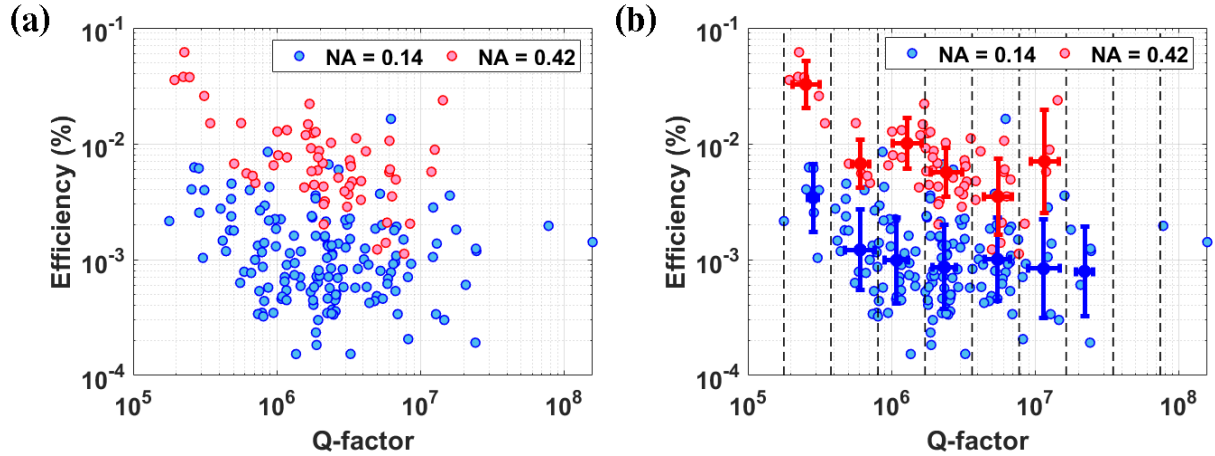

**Fig. S7** Efficiency vs Q-factor from two objective lenses: NA of 0.14 and 0.42. **(a)** Raw data. **(b)** Data is divided into 10 groups by Q-factor. Dashed black lines indicate the boundary of each group. Average Q-factor and % efficiency are plotted as a single point on each group with standard deviation as an error bar.

Fig. S7(a) shows the percent efficiency and Q-factor of 140 resonance modes from 57 microtoroids with two different objective lenses: NA of 0.14 and 0.42. To create this plot, we divided the data into ten groups by Q-factor. The dashed black lines indicate the boundary of each group, which is equally separated by Q-factor in log-scale. The average Q-factor and percent efficiency in each group is represented as a single point. Error bars represent the standard deviation in each group as shown by solid lines in Fig. S7(b).

## Supplementary Note 4: Additional coupling map

To observe the modification of line shape as the beam-microtoroid distance changes, a two-dimensional scan in the YZ plane is performed. In addition to Fig. 3, Fig. S8(a-b) shows how the resonance line shape changes when scanning the along the z-axis. We defined $(y, z)$ to be $(0,0)$ at the highest efficiency position. The asymmetric profiles of the changing Fano line shape imply a change in phase difference between two interfering modes.

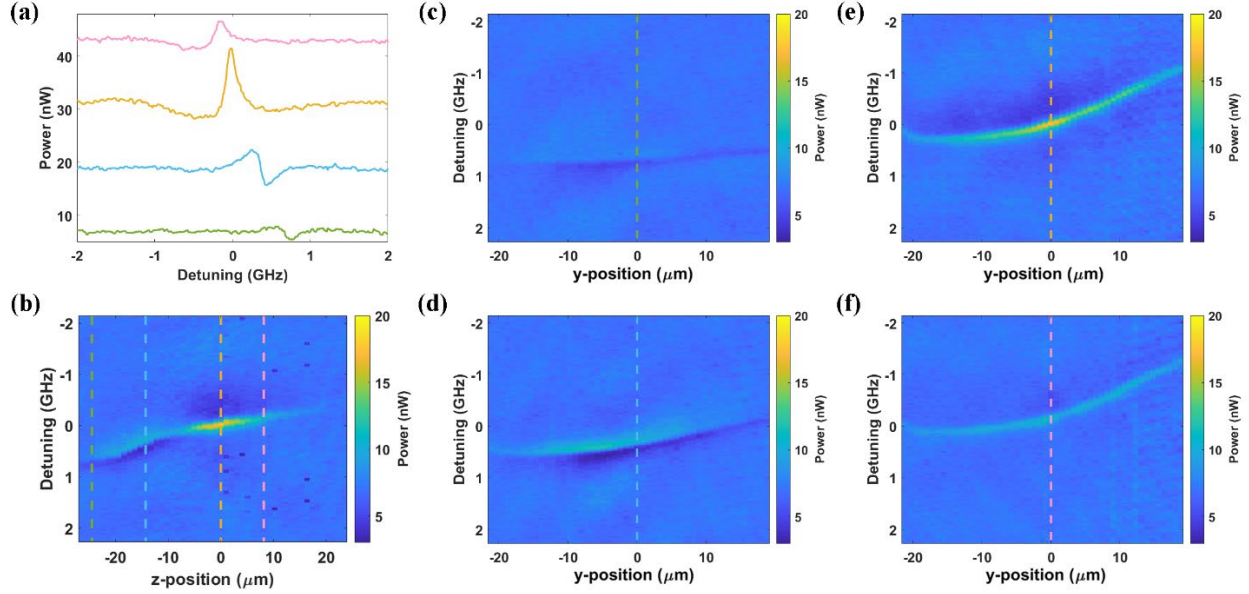

**Fig. S8** Mode spectra using a 5× objective lens (NA = 0.14) at different beam-microtoroid distances. **(a)** From bottom to top: the microtoroid was moved upward. Effectively, the beam was moved downward relative to the cavity. Spectra colors extracted from different z-positons correspond to the mode spectra in (b). **(b)** Mode spectra scanning along the z-axis at y = 0. **(c-f)** Mode spectra scanning along the y-axis at different z-positions. The same color of the dashed line means the same spectra shown in (a) and (b).

In addition to Fig. 5, Fig. S9(a) shows the resonance profile transition at different z-positions. There are at least four modes in this profile. Eq. (S8) was modified to be a summation of four terms to be used for fitting. Since changing the microtoroid position affects the change of effective index differently for different modes, the mode interaction is different at different positions, producing different profiles.

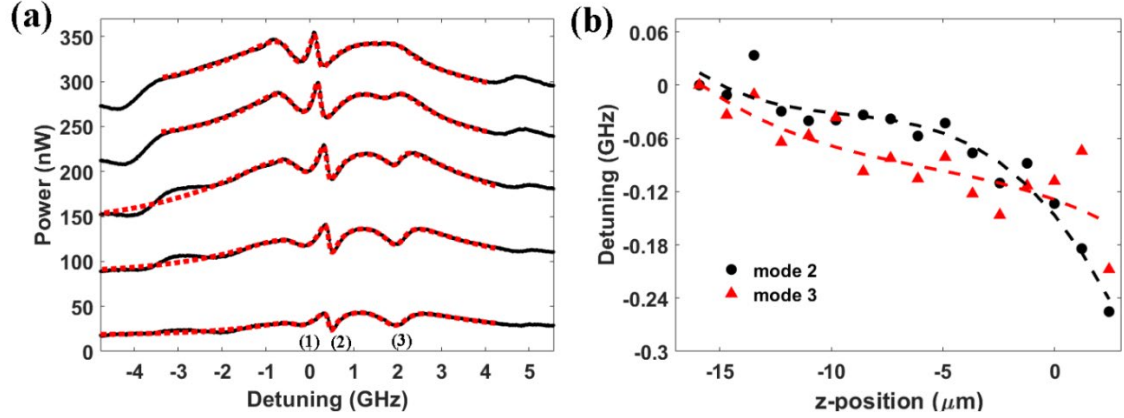

**Fig. S9** Mode spectra using a 5× objective lens (NA = 0.14) at different beam-microtoroid distances. **(a)** From bottom to top: the microtoroid was moved upward. Effectively, the beam was moved downward relative to the cavity. Resonance profiles are fitted with the summation of four modes. The numbers indicate modes. The 4$^{th}$ mode is a low Q-factor mode. **(b)** Resonance shift at different z-positions of two modes: (2) and (3).

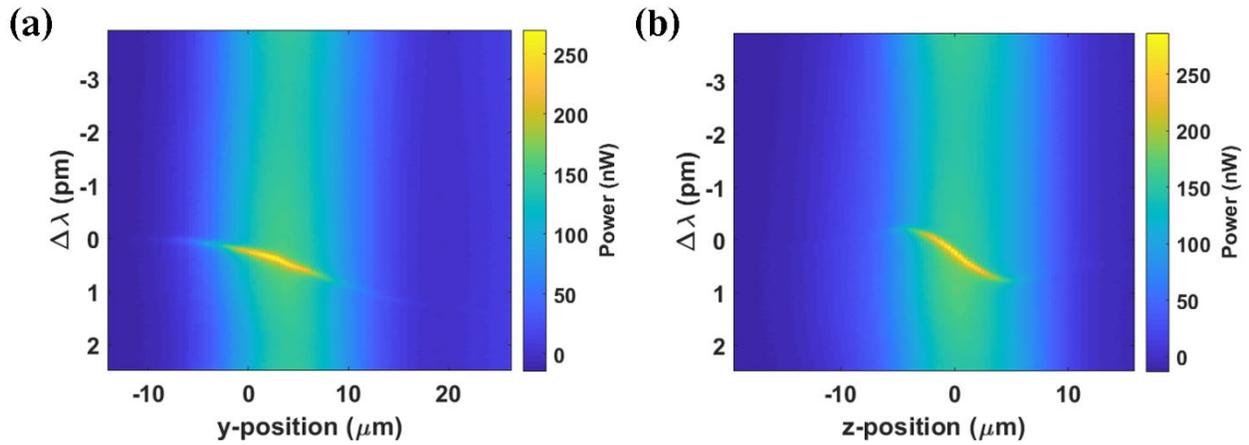

**Fig. S10** Pre-analyzed mode spectra at different positions using a 5x objective lens (NA = 0.14) **(a)** Detuning ($\Delta\lambda$) vs. y-position. **(b)** Detunning ($\Delta\lambda$) vs. z-position

Fig. S10 shows the pre-analyzed data of Fig. 4(c-d). In Fig. S10, it is apparent that when a resonance mode has higher power, the background light is also greater. There are two main sources of background light: resonant scattered light from low Q-modes and stray light. To filter out low Q-modes, we subtracted the average background at each position. This is shown in Fig. 4(c-d).

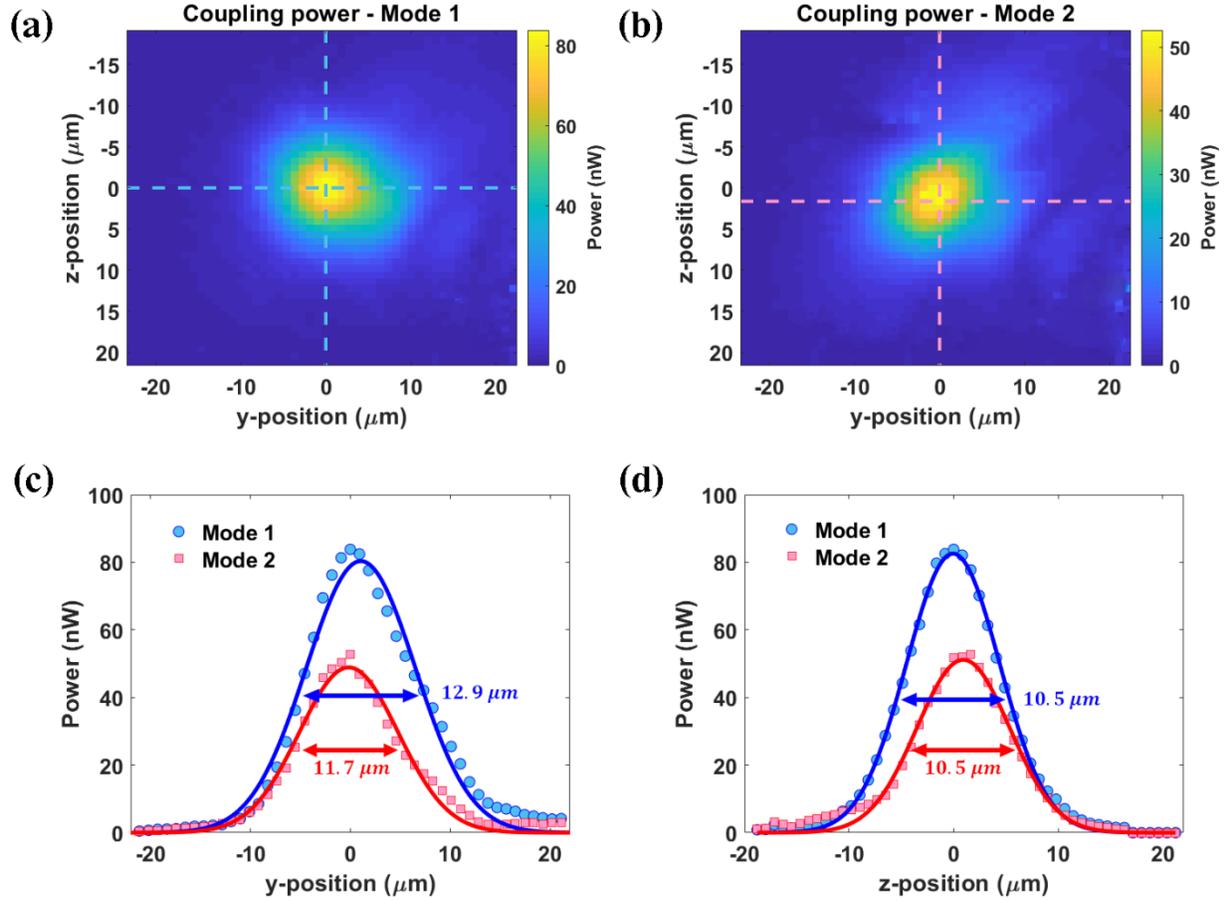

**Fig. S11** Coupling maps of the resonance curve shown in Fig. 5(a): (a) mode 1 and (b) mode 2. (y,z) = (0,0) is defined at the highest power of mode 1. **(a)** Coupling map of mode 1. The dashed blue lines along the y-axis and z-axis cross at (0,0) **(b)** Coupling map of mode 2. The dashed pink lines along the y-axis and z-axis cross at the highest power of mode 2. **(c)** Coupling power along y-axis. Points show power data along the vertical dashed blue line in (a) for mode 1 and power along vertical dashed pink line in (b) for mode 2. **(d)** Coupling power along the z-axis. Points show power data along the horizontal dashed blue line in (a) for mode 1 and power along the horizontal dashed pink line in (b) for mode 2. Solid lines are fit with a Gaussian equation.

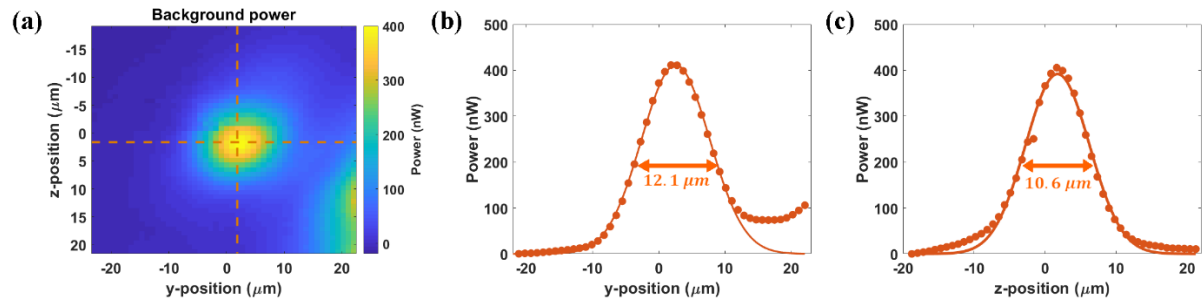

**Fig. S12** Background power map of results in Fig. 5 and Fig. S11. **(a)** Background power map. Dashed orange lines cross at the maximum background power. **(b)** Coupling power along the y-axis. The points show power data along the vertical dashed orange line in (a). **(c)** Coupling power along the z-axis. Points show power data along the horizontal dashed orange line in (a). Solid lines are fits to a Gaussian equation.

In addition to Fig. 5, resonance curves with a generalized Fano line shape were fit with Eq. (S8). Coupling maps of both modes (Fig. S11(a-b)) were generated by calculating mode power using Eq. (S9). The maximum power position is slightly different between the two modes. By fitting coupling power along the *y*-axis and *z*-axis to Gaussian equations, the FWHM of both modes, ~10 μm, were obtained as shown in Fig. S11(c-d). The background power, Fig. S12, is the $B$ parameter achieved by fitting Eq. (S9) at each position.

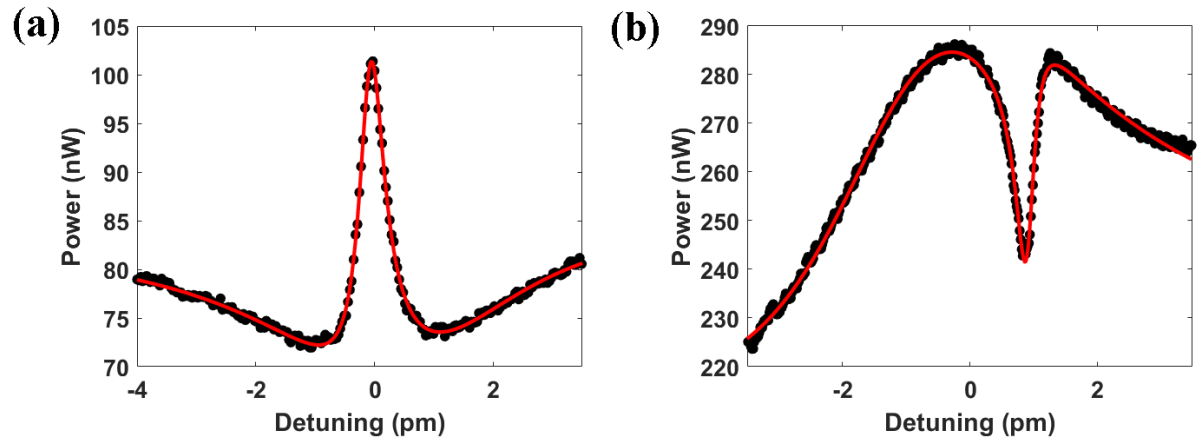

**Fig. S13** Resonance curves using a 20x objective lens (NA = 0.42).

For comparison, we replaced the 5× objective lens with a 20× objective. Fig. S13(a) and (b) show the resonance profiles at (y, z) = (0,0) of Fig. 6(a) and (b), respectively.

**Supplementary Note 5:** Finite element simulation

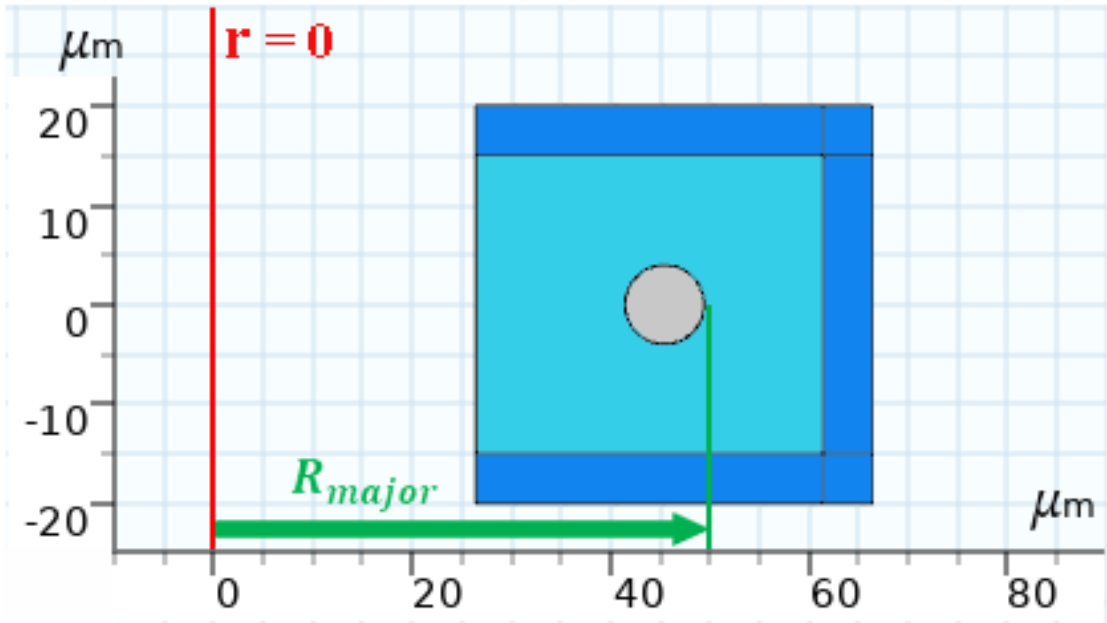

**Fig. S14** Finite element simulation geometry for computing electric field in microtoroid cavity. The red line is the azimuthal axis. The grey area is the microtoroid cavity whose refractive index is 1.45. The surrounding blue area is air. The dark blue area is a Perfectly Matched Layer.

A finite element simulation using COMSOL Multiphysics was performed to simulate the electric field in the microtoroid cavity. The system geometry was created in two dimensions around an azimuthal axis as shown in Fig. S14. The simulation was performed by using the Electromagnetic Waves module in the frequency domain. The microtoroid ring, indicated by the grey circle, was made of silica whose refractive index was set to be $n_{SiO_2} = 1.45$. Based on dimensions obtained from SEM images (see Supplementary Note 1), the major radius ($R_{major}$) and minor radius ($R_{minor}$) were 50 µm and 4 µm, respectively. We studied operation in air. Therefore, the blue and dark blue areas had a refractive index of 1. The dark blue area was defined to be a Perfectly Matched Layer. As an initial guess, the azimuthal mode number was swept $\pm 10$ around $\left\lfloor \frac{2\pi R_{major}}{n_{SiO_2}} \right\rfloor$ with step size of 1, with an approximate eigenfrequency guess of $\frac{c}{n_{SiO_2}\lambda}$, where $c$ and $\lambda$ are the speed of light and approximate operating wavelength (780 nm). The azimuthal mode number that yielded the relevant eigenfrequency within the tunable laser range was then selected.